# Transport characteristics of type II Weyl semimetal MoTe$_2$ thin films grown by chemical vapor deposition

Niraj Bhattarai[1, 2 *], Andrew W. Forbes[1, 2], Rajendra P. Dulal[3], Ian L. Pegg[1, 2], and John Philip[1, 2]

[1]Department of Physics, The Catholic University of America, Washington, DC 20064;

[2]The Vitreous State Laboratory, The Catholic University of America, Washington, DC 20064;

[3]Advanced Physics Laboratory, Institute for Quantum Studies, Chapman University, Burtonsville, MD, 20806

[*]Corresponding author: bhattarai@cua.edu

**Abstract**

Theoretical calculations and experimental observations show MoTe$_2$ is a type II Weyl semimetal, along with many members of transition metal dichalcogenides family. We have grown highly crystalline large-area MoTe$_2$ thin films on Si/SiO$_2$ substrates by chemical vapor deposition. Very uniform, continuous, and smooth films were obtained as confirmed by scanning electron microscopy and atomic force microscopy analyses. Measurements of the temperature dependence of longitudinal resistivity and current-voltage characteristics at different temperature are discussed. Unsaturated, positive quadratic magnetoresistance of the as-grown thin films has been observed from 10 K to 200 K. Hall resistivity measurements confirm the majority charge carriers are hole.

## A. INTRODUCTION

Transition metal dichalcogenides (TMDs) are the interesting class of materials that have been extensively studied in recent decades due to their diverse structural and electronic properties. [1-4]. TMDs are chemically represented by MX$_2$, where M is one of the transition metal atoms, and X is a chalcogen atom (S, Se, Te) stacked vertically one over the other with a weak van der Waals interaction between the successive planes [5, 6]. Their diverse physical properties make them excellent candidates for various technological applications, for example, in electronics, spintronics, optics, biosensors, catalysis, and magnetic sensors [4, 7-15]. One of the fascinating members of TMD family, MoTe$_2$ offers a wide possibility for many novel device applications as it can exist in different phases: hexagonal α-phase (2H) and monoclinic β-phase (1T$^{'}$) [16, 17]. The monoclinic 1T$^{'}$-MoTe$_2$ is a semimetal which undergoes a structural transition into an orthorhombic γ- phase, frequently referred to as the T$_d$ phase, at around 240 K [18-21]. Recent theoretical and an experimental study based on angle-resolved photoemission spectroscopy have revealed that MoTe$_2$ hosts type II Weyl semimetal fermions [19, 22]. T$_d$-MoTe$_2$ belongs to a non-centrosymmetric (inversion symmetry breaking) space group *Pmn*2$_1$ and is the only phase of MoTe$_2$ which hosts Weyl fermions. These Weyl fermions emerge at the topologically protected touching points between electron and hole pockets [19]. The Weyl semimetal behavior in materials requires either time-reversal symmetry or crystal inversion symmetry, or both [23]. Type II Weyl semimetals are known for exhibiting exotic low energy physics such as Fermi arcs on the surface, an anomalous Hall effect, distinct magneto-transport properties, and chiral anomaly induced quantum transport [24-27]. Bulk $MoTe_2$ has been recently reported to exhibit pressure-driven superconductivity, which has potential applications in quantum computation [2]. Furthermore, an extremely large positive magnetoresistance has also been reported in bulk as well as a monolayer of $MoTe_2$ [13, 28]. Two to four pairs of Weyl points in the $k_z = 0$ plane of the Brillouin zone of $MoTe_2$ were predicted [29, 30]. Fermi arcs from the projections of these Weyl points yield different scenarios. Experimentally different pictures of Fermi arcs are reported [19, 31, 32, 33], and it is interesting to explore large-area films to understand the transport behavior which is closely connected to the Weyl points and Fermi arcs. Although there has been enormous research on CVD grown $MoTe_2$, with our best knowledge, there are no reports on CVD grown thin films thicker than mono- and few-layer $MoTe_2$ [3, 5, 34, 35]. The $1T'$-$MoTe_2$ is predicted to be a large-gap spin-Hall insulator, and may find applications in spintronic devices and quantum computation [3]. For such applications, large-area growth of the material is essential. Growing large area $MoTe_2$ thin films is significantly more challenging than growing small flakes of 2D $MoTe_2$. Particularly with CVD method, the physical conditions play a great role in achieving single phase films. Maintaining a right ratio of precursors, adequate flow of carrier gases and temperature controlling during growth is also critical for yielding uniform film over a large area of the substrates.

In this article, we report the detailed growth, electrical, and magneto-transport properties of ~30 nm thick $MoTe_2$ films grown by CVD. We obtained a near stoichiometric monoclinic phase of highly crystalline $MoTe_2$ thin films distributed uniformly over a large area of $Si/SiO_2$ wafer. We studied the structural and morphological characteristics of the as-grown samples, and carried out extensive transport measurements. Our experimental results on electrical transport measurements have shown that these films are semimetallic and display a positive non-saturating magnetoresistance at different temperatures.

## B. RESULTS AND DISCUSSION

Large area $MoTe_2$ films were grown by CVD method. Figure 1 show the x-ray diffraction (XRD) patterns of thin films obtained from CVD growth with different combinations of the amount of precursor materials, carrier gas flow, and heating protocol. The detailed experimental procedure for the growth, fabrication of devices, and characterization of the thin films can be found in the experimental section. In Figure 1, the first three XRD patterns (a, b, and c) are from trials that yielded relatively low quality of films, while curve (d) shows the XRD peaks of a successfully grown high quality MoTe$_2$ film. In order to get the best estimates of the structural parameters, a whole pattern fitting Rietveld refinement of the peaks was performed in the range $10° \leq 2\theta \leq 67°$ (represented by green curve in Figure 2 (a)). The refinement was consistent with a monoclinic (1T$'$- MoTe$_2$) crystal structure of the centrosymmetric space group $P2_1/m$ (11) with lattice constants $a = 6.34$ Å, $b = 3.45$ Å and $c = 13.85$ Å. the best fit lattice constants are in excellent agreement with previously reported data [28, 36]. Strong diffraction peaks were observed near 13°, 26°, 39°, 53°, and 66°. In the monoclinic crystal structure $P2_1/m$ (11), those diffraction peaks correspond to Miller indices (002), (004), (006), (008), and ($\bar{2}24$) respectively [19, 28].

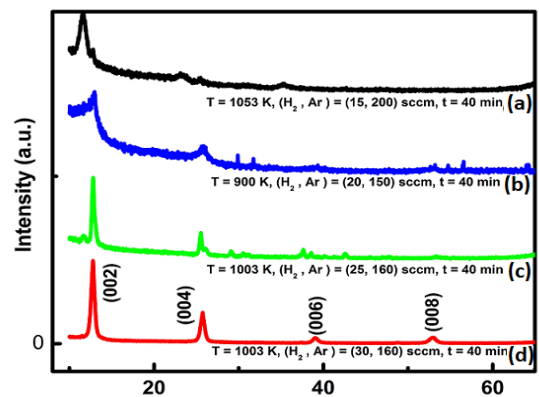

**Figure 1:** A comparative plot of XRD patterns of CVD grown $MoTe_2$ thin film samples grown for 40 minutes at different physical conditions. The precursor combination is different for curve (a) and rest of the cases.

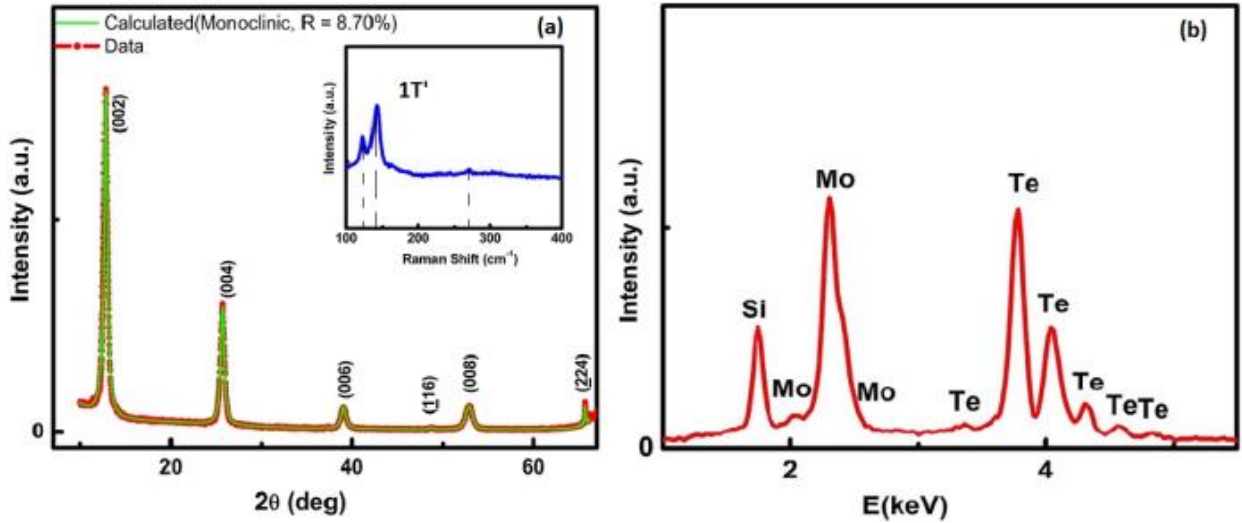

**Figure 2:** (a) X-ray diffraction data of MoTe$_2$ thin film grown on Si/SiO$_2$ substrates with Rietveld refinement. Data are refined in the range $10° \leq 2\theta \leq 67°$ and the best fit was obtained with a discrepancy index R of 8.70%. The inset shows the Raman spectroscopy measurement of the MoTe$_2$ thin film. The Raman peaks are observed near 121, 141, and 270 cm$^{-1}$ (1T$'$). (b) Energy dispersive x-ray spectroscopy spectrum of MoTe$_2$ thin film.

The results from the x-ray spectroscopy were further supported by the Raman spectroscopy measurements of the samples. The Raman spectrum of MoTe$_2$ thin film is displayed in the inset which shows peaks at 121, 141, and 270 cm$^{-1}$ which are the representative Raman peaks of 1T$'$-MoTe$_2$ thin films [16]. Qualitative elemental analysis was performed using energy dispersive spectroscopy (EDS) as shown in Figure 2 (b). Multiple spot scanning of the sample confirmed the average chemical ratio of molybdenum and tellurium was very close to the expected stoichiometric ratio such that Mo:Te ~ 1:2.

The scanning electron microscopy (SEM) image in Figure 3 (a) and the atomic force microscopy (AFM) image in Figure 3 (b) show the morphology and topography of the MoTe$_2$ thin films grown by CVD. We observed continuous thin film distributed very uniformly over a large area of the silicon substrates. We were able to synthesize uniform and continuous films in dimension up to 4 mm × 8 mm (an optical image of large area film is shown as an inset in the upper left corner of Figure 3(a)). The inset in the upper left corner of Figure 3 (b) displays the height profile of a film. As displayed, the height profile of the film was measured using AFM along the blue line drawn horizontally starting from a part of the bare substrate into the film surface. A typical 40 minutes of growth yielded films of 31 nm of average thickness. The thickness of the films was observed depending on the duration of the growth. In general, longer growth time yielded thicker films. The topography of the surface is consistent across different parts of the substrate and the root mean squared average roughness of the film surface measured by AFM was less than 7 nm.

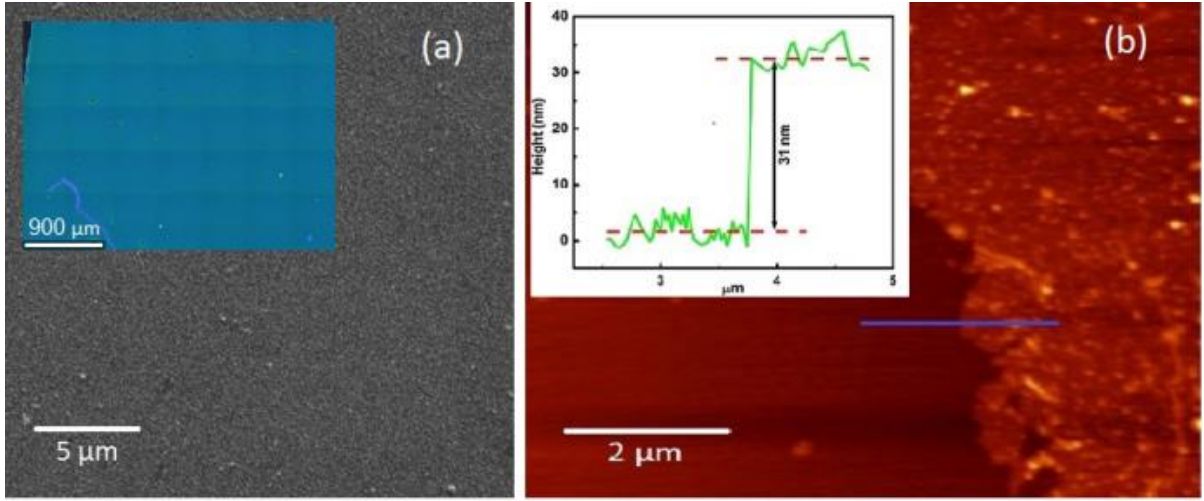

**Figure 3:** Morphology of MoTe$_2$ thin films. (a) Scanning electron microscopy image. Inset shows the large area optical image taken by multiple image stitching method in Raman microscopy. (b) Atomic force microscopy image and corresponding height profile as inset in the upper left corner. The thickness of the thin film is measured along the blue line.

As discussed in the experimental section, thin film devices of MoTe$_2$ were fabricated using gold wires as electrodes attached on the thin film surface using indium. The uniformly distributed large area MoTe$_2$ thin films greatly facilitated the fabrication of electrical devices for electronic measurements. Figure 4 (a) shows the temperature dependence of the longitudinal resistivity ($\rho_{xx}$) measured between 5 and 300 K. The resistivity increased with increasing temperature showing a residual resistivity ratio (RRR) $\rho_{xx}(300\ \text{K}) / \rho_{xx}(5\ \text{K}) = 4.2$. This value is comparable with previous reports on similar phases of MoTe$_2$ [2, 20, 21]. One study on bulk 1T'- MoTe$_2$ grown by flux methods reports a large RRR greater by a factor of hundred compared to our result [13]. This difference in RRR may be assigned to the varied mechanism used for growing the samples plus the different thickness of the samples used for measurements in contrast to ours. At 5 K, the resistivity is $212 \times 10^{-6}$ Ω cm, which increases to $893 \times 10^{-6}$ Ω cm at room temperature. The room temperature resistivity in our sample is very much consistent with literature [13]. The inset in the lower-right corner displays a fit to the resistivity data in between 5 and 35 K to a combination of electron-phonon scattering and Fermi-liquid mechanisms. The form of the fitted equation is $\rho_{xx} = \rho_0 + cT^2 + gT^5$ where $\rho_0$ is the residual resistivity due to impurity scattering; $c$ and $g$ are fitting parameters to Fermi-liquid ($\sim T^2$) and electron-phonon scattering ($\sim T^5$) mechanisms at low temperatures [13, 23]. From the best fit, we found $\rho_0 = 210$ μΩ cm and other fitting parameters are $c = 711 \times 10^{-4}$ μΩ cm K$^{-2}$, and $g = 212 \times 10^{-10}$ μΩ cm K$^{-5}$. The small value of $g$ hints substantial phonon drag in our system, and trivial value of $c$ indicates weak electron-electron scattering. The inset in the upper left corner is the magnified plot of resistivity in between 200 to 300 K. The change in resistivity at temperature 250 K indicates the structural transition from room temperature monoclinic (1T$'$-MoTe$_2$) phase to orthorhombic (T$_d$ -MoTe$_2$) phase at low temperatures, consistent with previous reports [1, 19, 20].

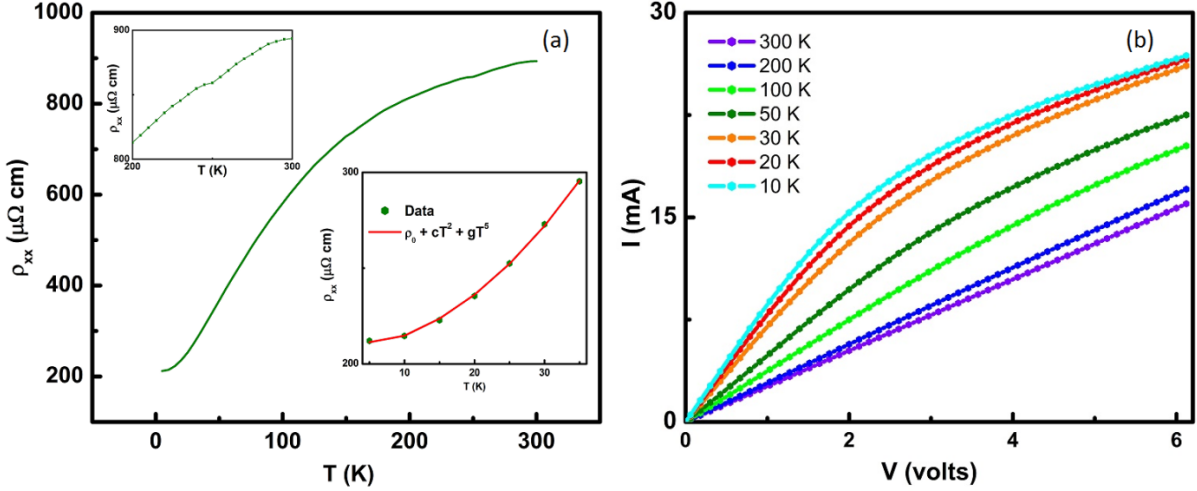

**Figure 4:** Transport measurements of MoTe$_2$ thin film devices. (a) Longitudinal resistivity measured between 5 and 300 K. The inset in the lower right corner is the fit of resistivity to a combination of electron phonon scattering and Fermi-liquid mechanisms, and the inset in the upper left corner shows a dent in resistivity curve indicating structural phase transition at 250 K. (b) Current – voltage characteristics measured at different temperatures.

In addition to longitudinal resistivity, we measured the current-voltage (I – V) characteristics of the MoTe$_2$ thin film devices at different temperatures. Figure 4 (b) shows the typical I–V characteristics measured from room temperature to 10 K. There is a steady increase in current with a decrease in temperature. As the temperature is lowered from room temperature to 10 K, there is a 70% increase in the current flowing through the thin film devices at 1 V bias voltage. Similarly, at 3 and 6 V bias voltages; there was a relative change in current by 59% and 40% concerning room temperature, and at 10 K. Hence, it was observed that the magnitude of the input bias voltage is critical to determine the efficiency of devices fabricated from the as-grown MoTe$_2$ thin films. The I – V curves measured at room temperature, and 200 K are Ohmic showing the metallic behavior of the thin film at those temperatures. But as we lower the temperature below 200 K, there is a slight deviation from the linearity starting at 100 K. The I – V curves below 100 K are observed to concave downward above 2 V bias voltages. Such non-linearity of I – V curves below 100 K may be the associated with the monoclinic-orthorhombic phase transition and can be the characteristics of T$_d$ phase Weyl semimetallic MoTe$_2$ [37].

Figure 5 displays the magnetoresistance as a function of the magnetic field applied parallel to the direction of current flow in the field range ±3 T measured in between 200 and 10 K. Magnetoresistance (MR) is expressed in percentage as MR = $\frac{\rho(T) - \rho(0)}{\rho(0)}$ × *100%,* where $\rho(T)$ and $\rho(0)$ are the resistivity calculated at magnetic field T and at zero field respectively. The magnitude of MR reaches 9.6% at 3 T and 10 K. Similarly at 3 T, MR of

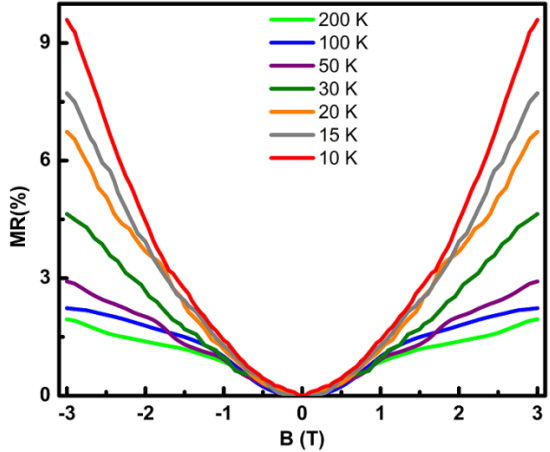

**Figure 5:** Magnetoresistance as a function of the magnetic field applied parallel to the direction of current flow.

7.71%, 6.73%, 4.64%, 2.90%, 2.25%, and 1.95% were observed at 15 K, 20 K, 30 K, 50 K, 100 K, and 200 K respectively. Resulting MR is of a few orders smaller than what has been reported for the $T_d$- MoTe$_2$ crystals [13, 21, 38]. MR is temperature dependent, and varies significantly with strength of the applied magnetic field. We observed a positive as well as unsaturated magnetoresistance for temperatures up to 200 K and a magnetic field up to 3 T and the MR decreases with the increasing temperature. The MR show a quadratic dependence on magnetic fields at lower fields, and tends to stabilize to linear for larger magnetic fields, which is in good agreement with recent reports on few layered as well as bulk MoTe$_2$ [21, 28]. Similar MR behavior also has been reported in many other well-known topological semimetallic materials, such as WTe$_2$, TaAs, and NiTe$_2$ [39-41]. Positive and unsaturated magnetoresistance in $T_d$- MoTe$_2$ may be attributed with the Lifshitz transition induced Fermi surface reconstruction. Such reconstruction results change the electronic structure and acts as a driving force for the electron-hole compensation effect [38]. However, the origin of MR in semimetals is still a subject of intense study.

We carried out Hall resistivity measurements at different temperatures starting from 300 K temperature to 10 K, as shown in Figure 6 (a). Hall resistivity measurements were performed in the field between ±3 T using a Hall bar shape device as shown as an inset in the upper left corner. The measured Hall resistivity has a strong linear dependence on magnetic field reflecting the dominance of one kind of charge carrier in electronic transport of our CVD grown MoTe$_2$ thin film samples. Such a linear dependence of the Hall resistivity to the magnetic field in a semimetal like MoTe$_2$ shows that a situation with charge carriers, both electrons and holes and are nearly compensated in our thin films [38]. The Hall resistivity exhibits a positive slope which indicates that electronic transport is dominated by holes, or that the holes have higher mobilities than the electrons. To estimate the mobility of the charge carriers, we have fitted a power law using MR = $bB^n$ to the MR curve at 10 K in Figure 6(b) [42]. For an equal number of electron density ($n_e$) and hole density ($n_p$) in a two band model, the MR is proportional to the square of the magnetic field as in the relation $\mu_e\mu_p B^2$, where $\mu_e$ and $\mu_p$ are the mobilities of electron and holes [42, 43]. Figure 6(b) shows a fit with n = 1.85, with a nearly compensated condition (for perfect compensation n = 2). Using the fit to the data (as shown in inset Figure 6(b)), we have calculated the geometric-mean mobility $\mu_m = \sqrt{\mu_e\mu_p}$ = ~$1.05 \times 10^4$ cm$^2$ V$^{-1}$ s$^{-1}$. This value is comparable to the results reported in $T_d$-MoTe$_2$ reported in previous literatures [13, 21, 28, 42]. However, our large-area films with thickness close to 30 nm show that the magnetoresistance is lower than that previously reported in monolayers and 3D MoTe$_2$ samples.

## C. SUMMARY AND CONCLUSIONS

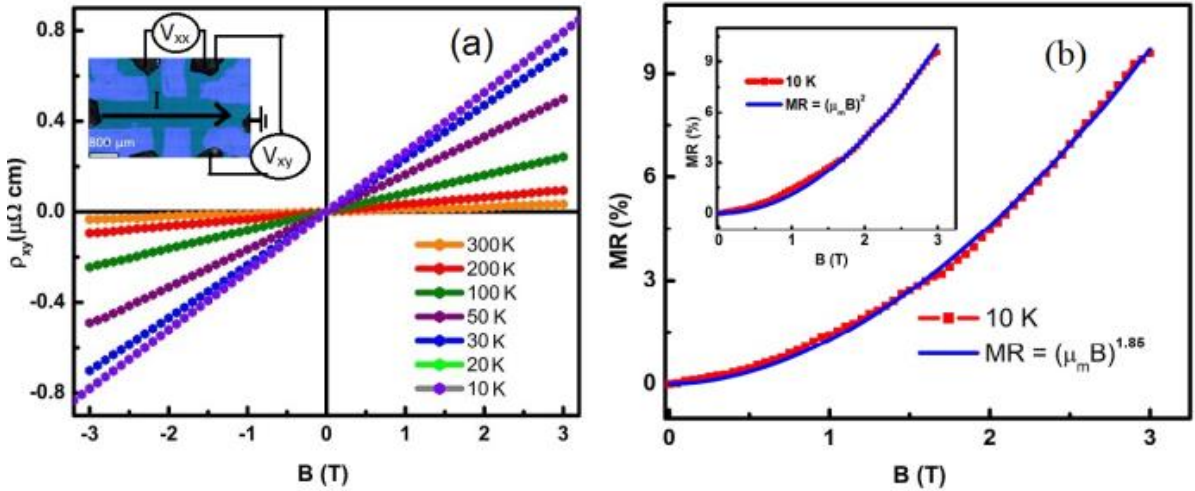

**Figure 6:** (a) Hall resistivity measurements carried out in between 10 and 300 K. Inset in the upper left corner shows the optical image of the Hall device used for the measurements. (b) Fit to the magnetoresistance data in Figure 5 at 10 K. The blue solid curve represents the fitted curve using the power law formula MR = $bB^n$. Inset shows the fitting curve using MR= $(\mu_m B)^2$.

We have successfully grown highly crystalline 31 nm thick large-area thin films of MoTe$_2$ by chemical vapor deposition method and found a monoclinic 1T'-phase structure with a space group $P2_1/m$ (11). We have discussed the procedure of large-area growth of continuous thin films and investigated the temperature dependence of longitudinal resistivity and I-V characteristics at different temperatures of as-grown MoTe$_2$ films. We observed the semimetallic behavior with holes dominating the majority of charge carriers. There is quadratic dependence of positive magnetoresistance at lower fields which tends to stabilize to linear in higher fields and remains unsaturated within the field range probed. These results are consistent with previous experimental results. However, the differences in MR may involve thickness-dependent effects and could use further investigation in the future. Monolayers to a few layers of MoTe$_2$ are of great interest for understanding the basic physics of Weyl semimetals. Our future study will be focused to obtain large-area MoTe$_2$ with smaller thicknesses. These investigations will lead to better understand large-area MoTe$_2$ thin films for future device applications.

## D. EXPERIMENTAL SECTION

MoTe$_2$ thin films were grown by chemical vapor deposition. The growth was carried out in a long quartz tube of an inch diameter placed in a temperature-controlled furnace. A mixture of high-purity argon (Ar) and hydrogen (H$_2$) gases were used as carrier gases, which forms a reducing environment during the CVD process. In the course of the CVD, a quartz boat containing the mixed powder of 0.20 g tellurium (Te), 0.12 g of molybdenum pentachloride (MoCl$_5$), and 0.12 g of molybdenum (VI) trioxide (MoO$_3$) was placed at the center of the furnace. The precursors used are of their highest quality purchased from Sigma Aldrich and Alfa Aesar. The purpose of mixing up the Te powder, with MoCl$_5$ and MoO$_3$ powder is to lower the melting point of the precursors during growth mechanism [35]. But as temperature of the center of the furnace is much higher than melting point of Te alone, that results in higher rate of losing Te. Also, the stoichiometry demands more Te compared to Mo, thus another quartz boat containing 0.25 g of Te powder was placed in the upstream such that the Te source temperature did not exceed 753 K. This additional upstream Te source helped to generate consistent and sufficient amount of Te vapor in the quartz tube which ultimately helps to yield high quality $MoTe_2$ films and avoiding deficiency of Te during the whole process. Cleaned silicon wafer pieces capped with 300 nm of silicon dioxide were used as substrates and were placed in the downstream of the tube on a quartz slide at a distance 3 – 6 cm from the center of the furnace. Silicon wafers used as substrates were sonicated for 3 minutes in warm deionized water and cleaned using isopropyl alcohol, acetone, and deionized water. Weighing the precursors was done in a glove box under nitrogen ambient while the tube was continuously purged with argon during the installment of the precursors in the quartz tube. The furnace was ramped up to 1003 K in 25 minutes and kept at that temperature for the next 40 minutes in the presence of 160 sccm Ar and 30 sccm $H_2$ gas flow. The furnace was allowed to cool naturally up to 623 K and then rapidly brought down to room temperature by lifting up the lid of the furnace and exposing the quartz tube in cold air. Quenching below 623 K is important as it helps to form the $1T^{'}$- $MoTe_2$ [18]. A schematic of the CVD setup and heating protocol used in our experiment is shown in Figure 7.

We have tailored the growth conditions to obtain high-quality large-area $MoTe_2$ films. The growth procedure is similar to what is reported in reference [35]. Figure 1 in the results and discussion section shows the XRD patterns of the films obtained from different CVD growths. Initially, we have grown according to the conditions outlined in reference [35] setting different durations of growth, but large-area films were obtained only with 40 or higher minutes of growth time. XRD pattern of a film prepared in 40 minutes of growth time using 0.50 g of Te in the upstream and mixed precursor of 30 g ($MoO_3$: $MoCl_5$: Te= 1:1:1) in the center of quartz tube, indicated mixed phases as shown in curve (a) of Figure 1. Then, we performed several other trials considering changes in heating protocol, the ratio of precursor materials, and carrier gas flow to obtain high quality single phase $MoTe_2$ films. Additional Te powder was added in the mixture to assist in lowering the melting point of precursors as long as the CVD run time. Figure 1, the curve (b) represents the XRD pattern of a typical synthesis of 40 minutes growth time at 900 K, in the presence of 150 sccm Ar and 20 sccm $H_2$ gas flow and with mixed precursor of 44 g ($MoO_3$, $MoCl_5$, and Te = 12, 12, and 20 g respectively). Similarly, the curve (c) represents the XRD pattern of a typical synthesis of 40 minutes growth time at temperature 1003 K and in the presence of 160 sccm Ar and 25 sccm $H_2$ gas flow. The quality of the films grown was improved greatly and was close to the expected stoichiometric atomic ratio, but multiple phases still appeared in the XRD patterns. Increasing the amount of carrier gas in these cases facilitated in the reaction process and transportation of the vapor phase precursors. Increasing H$_2$ helped in creating a more reducing atmosphere, enhancing the reaction between reduced Mo and Te vapor. High-quality large-area films were obtained only when the H$_2$ concentration in the gas mixture was increased to 30 sccm. The curve (d) as shown in Figure 1 is the XRD pattern of as-grown high-quality films whose detailed crystallographic analysis is explained in section results and discussion. Generally, the film thickness was observed to increases with the increase in duration of the growth.

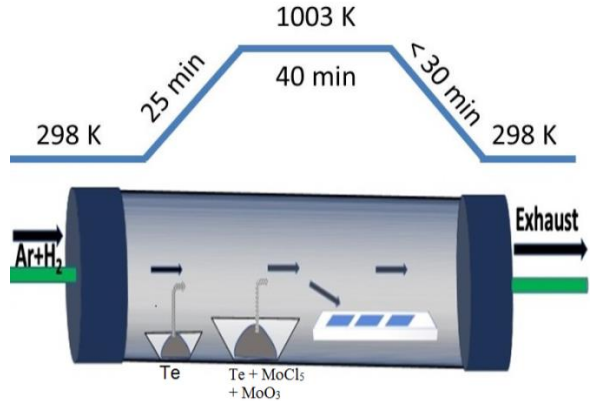

**Figure 7:** Schematic of chemical vapor deposition method and the heating protocol used.

## E. CHARACTERIZATION

The morphology of as-grown thin films was analyzed using the scanning electron microscopy [JEOL JSM-5910LV]. The surface topography, film thickness and Raman spectroscopy mapping were done using Raman AFM system [Alpha 300 RA, WITec]. The crystal structure of the samples was investigated by using an X-ray diffractometer (Thermo / ARL X'TRA) with Cu-$K_\alpha$ radiation ($\lambda$= 1.54059 Å) [44]; structural parameters were estimated by a full Rietveld refinement using Materials Data Inc. JADE 9 software; composition by energy dispersive spectroscopy using Oxford instruments Ultim Max with 170 $mm^2$ aperture coupled with SEM [JEOL JSM-5910LV]; the electrical transport measurements were carried out using a Quantum Design physical property measurement system (PPMS) AC transport with horizontal rotator option. The longitudinal resistivity and magnetoresistance measurements were done in standard four-probe geometry, whereas Hall measurements were carried out using Hall bar. To avoid possible doping of the film during lithographic processing, we followed sharp metal scratch method. The film was scratched into a Hall bar shape manually using a sharp metal tip and Ohmic contacts were made using indium metal on each terminal. Previously, Hall shape devices fabricated by the metal scratch method have been used to measure the anomalous Hall effect in thin films. Such method helps to avoid possible doping of the film through lithographic processing [45, 46]. Current-voltage characteristics were measured using Keithley 2400 Source meter coupled with PPMS. The samples were connected to gold wires using indium contacts.

## Acknowledgment

This work is supported by The Vitreous State Laboratory.

## REFERENCES

1. W. G. Dawson, and D. W. Bullett: *Electronic structure and crystallography of MoTe$_2$ and WTe$_2$*. *Journal of Physics C: Solid State Physics* **20**(36), 6159 (1987).
2. Q. Yanpeng, P. G. Naumov, M. N. Ali, C. R. Rajamathi, W. Schnelle, O. Barkalov, M. Hanfland, S-C. Wu, C. Shekher, Y. Sun, V. Süß, M. Schmidt, U. Schwarz, E. Pippel, P.Werner, R. Hillebrand, T. Forster, E. Kampert, S. Parkin, R. J. Cava, C. Felser, B. Yan, and S. A. Medvedev: *Superconductivity in Weyl semimetal candidate MoTe$_2$*. *Nature communications* **7**, 11038 (2016).
3. C. H. Naylor, W. M. Parkin, J. Ping, Z. Gao, Y. R. Zhou, Y. Kim, F. Streller, R. W. Carpick, A. M. Rappe, M. Drndić, J. M. Kikkawa, and A. T. C. Johnson: *Monolayer single-crystal 1T'-MoTe$_2$ grown by chemical vapor deposition exhibits weak antilocalization effect*. *Nano letters* **16**(7), 4297 (2016).
4. X. Xu, W. Yao, D. Xiao, and T. F. Heniz: *Spin and pseudospins in layered transition metal dichalcogenides*. *Nature Physics* **10**(5), 343(2014).
5. L. Zhou, K. Xu, A. Zubair, X. Zhang, F. Ouyang, T. Palacios, M. S. Dresselhaus, Y. Li, and J. Kong: *Role of Molecular Sieves in the CVD Synthesis of Large-Area 2D MoTe$_2$*. *Advanced Functional Materials* **27**(3),1603491 (2017).

6. V. Nicolosi, M. Chhowalla, M. G. Kanatzidis, M. S. Strano, and J. N. Coleman: *Liquid exfoliation of layered materials*. *Science* **340**(6139), 1226419 (2013).
7. S. Manzeli, D. Ovchinnikov, D. Pasquier, O. V. Yazyev, and A. Kis: *2D transition metal dichalcogenides*. *Nature Reviews Materials* **2**(8), 17033 (2017).
8. S. H. Noh, J. Hwang, J. kang, M. H. Seo, D. Choi, and B. Han: *Tuning the catalytic activity of heterogeneous two-dimensional transition metal dichalcogenides for hydrogen evolution*. *Journal of Materials Chemistry A* **6**(41), 20005(2018).
9. C. Tan, and H. Zhang: *Two-dimensional transition metal dichalcogenide nanosheet-based composites. Chemical Society Reviews* **44**(9), 2713(2015).
10. X. Huang, Z. Zeng, and H. Zhang: *Metal dichalcogenide nanosheets: preparation, properties and applications*. *Chemical Society Reviews* **42**(5), 1934(2013).
11. Z. Yin, H. Li, H. Li, L. Jiang, Y. Shi, Y. Sun, G. Lu, Q. Zhang, X. Chen, and H. Zhang: *Single-layer $MoS_2$ phototransistors*. *ACS nano* **6**(1), 74(2011).
12. X. Huang, Z. Zeng, S. Bao, M. Wang, X. Qi, Z. Fan, and H. Zhang: *Solution-phase epitaxial growth of noble metal nanostructures on dispersible single-layer molybdenum disulfide nanosheets. Nature communications* **4**, 1444 (2013).
13. Q. Zhou, D. Rhodes, Q. R. Zhang, S. Tang, R. Schönemann, and L. Balicas: *Hall effect within the colossal magnetoresistive semimetallic state of $MoTe_2$*. *Physical Review B* **94**(12), 121101(2016).
14. X. Qian, J. Liu, L. Fu, and J. Li: *Quantum spin Hall effect in two- dimensional transition metal dichalcogenides*. *Science* **346**(6215), 1344 (2014).
15. R. P. Dulal, B. R. Dahal, A. Forbes, N. Bhattarai, I. L. Pegg, and J. Philip: *Nanostructures of type-II topological Dirac semimetal $NiTe_2$. Journal of Vacuum Science & Technology B* **37**(4), 042903 (2019).
16. S. Song, D. H. Keum, S. Cho, D. Perello, Y. Kim, and Y. H. Lee: *Room temperature semiconductor–metal transition of $MoTe_2$ thin films engineered by strain*. *Nano letters* **16(**1), 188(2015).
17. E. Revolinsky, and D. J. Beerntsen: *Electrical properties of α-and β-$MoTe_2$ as affected by stoichiometry and preparation temperature*. *Journal of Physics and Chemistry of Solids* **27**(3), 523(1966).
18. T. A. Empante, Y. Zhou, V. Klee, A. E. Nguyen, I-H. Lu, M. D. Valentin, S. A. N. Alvillar, E. Preciado, A. J. Berges, C. S. Merida, M. Gomez, S. Bobek, M. Isarraraz, E. J. Reed, and L. Bartels: *Chemical vapor deposition growth of few-layer $MoTe_2$ in the 2H, 1T', and 1T phases: tunable properties of $MoTe_2$ films*. *ACS nano* **11**(1), 905(2017).
19. K. Deng, G. Wan, P. Deng, K. Zhang, S. Ding, E. Wang, M. Yan, H. Huang, H. Zhang, Z. Xu, J. Denlinger, A. Fedorov, H. Yang, W. Duan, H. Yao, Y. Wu, S. Fan, H. Zhang, X. Chen, and S. Zhou: *Experimental observation of topological Fermi arcs in type-II Weyl semimetal $MoTe_2$*. *Nature Physics* **12**(12), 1105(2016).
20. T. Zandt, H. Dwelk, C. Janowitz, and R. Manzke: *Quadratic temperature dependence up to 50 K of the resistivity of metallic $MoTe_2$*. *Journal of alloys and compounds* **442**(1), 216(2007).
21. X. Luo, F. C. Chen, J.L. Zhang, Q. L. Pei, G. T. Lin, W. J. Lu, Y. Y. Han, C. Y. Xi, W. H. Song, and Y. P. Sun: *$T_d$-$MoTe_2$: A possible topological superconductor*. *Applied Physics Letters* **109**(10), 102601(2016).
22. A. A. Soluyanov, D. Gresch, Z. Wang, Q. S. Wu, M. Troyer, X. Dai, and B. A. Bernevig: *Type-II weyl semimetals*. *Nature* **527**(7579), 495(2015).
23. R. P. Dulal, B. R. Dahal, A. Forbes, N. Bhattarai, I. L. Pegg, and J. Philip: *Weak localization and small anomalous Hall conductivity in ferromagnetic Weyl semimetal $Co_2TiGe$*. *Scientific reports* **9**(1), 3342(2019).
24. X. Huang, L. Zhao, Y. Long, P. Wang, D. Chen, Z. Yang, H. Liang, M. Xue, H. Weng, Z. Fang, X. Dai, and G. Chen: *Observation of the chiral-anomaly-induced negative magnetoresistance in 3D Weyl semimetal TaAs*. *Physical Review X* **5**(3), 031023 (2015).
25. G. Xu, H. Weng, Z. Wang, X. Dai, and, Z. Fang: *Chern semimetal and the quantized anomalous Hall effect in $HgCr_2Se_4$*. *Physical review letters* **107**(18), 186806(2011).
26. Z. Wan, A. M. Turner, A. Vishwanath, and S. Y. Savrasov: *Topological semimetal and Fermi-arc surface states in the electronic structure of pyrochlore iridates*. *Physical Review B* **83**(20), 205101(2011).
27. S. A. Parameswaran, T. Grover, D. A. Abanin, D. A. Pesin, and A. Vishwanath: *Probing the chiral anomaly with nonlocal transport in three-dimensional topological semimetals*. *Physical Review X* **4**(3), 031035(2014).

28. D. H. Keum, S. Cho, J. H. Kim, D. H, Choe, H. J. Sung, M. Kan, H. Kang, J. Y. Hwang, S. W. Kim, H. Yang, K. J. Chang, and Y. H. Lee: *Bandgap opening in few-layered monoclinic MoTe$_2$*. Nature Physics **11**(6), 482(2015).
29. Y. Sun, S-C. Wu, M. N. Ali, C. Felser, and B. Yan: *Prediction of Weyl semimetal in orthorhombic MoTe$_2$*. Physical Review B **92**(16), 161107(2015).
30. Z. Wang, D. Gresch, A. A. Soluyanov, W. Xei, S. Kushwaha, X. Dai, M. Troyer, R. J. Cava, and B. A. Bernevig: *MoTe$_2$: a type-II Weyl topological metal*. Physical review letters **117**(5), 056805(2016).
31. L. Huang, T. M. McCormick, M. Ochi, Z. Zhao, M-T. Suzuki, R. Arita, Y. Wu, D. Mou, H. Cao, J. Yan, N. Trivedi, and A. Kaminski: *Spectroscopic evidence for a type II Weyl semimetallic state in MoTe$_2$*. Nature materials **15**(11), 1155(2016).
32. J. Jiang, Z. K. Liu, Y. Sun, H. F. Yang, C. R. Rajamathi, Y. P. Qi, L. X. Yang, C. Chen, H. Peng, C-C. Hwang, S. Z. Sun, S-K. Mo, I. Vobornik, J. Fujii, S. S. P. Parkin, C. Felser, B. H. Yan, and Y. L. Chen: *Signature of type-II Weyl semimetal phase in MoTe$_2$*. Nature communications **8**, 13973(2017).
33. A. Liang, J. Huang, S. Nie, Y. Ding, Q. Gao, C. Hu, S. He, Y. Zhang, C. Wang, B. Shen, J. Liu, P. Ai, L. Yu, X. Sun, W. Zhao, S. Lv, D. Liu, C. Li, Y. Zhang, Y. Hu, Y. Xu, L. Zhao, G. Liu, Z. Mao, X. Jia, F. Zhang, S. Zhang, F. Yang, Z. Wang, Q. Peng, H. Weng, X. Dai, Z. Fang, Z. Xu, C. Chen, and X. J. Zhou: *Electronic evidence for type II Weyl semimetal state in MoTe$_2$*. arXiv preprint arXiv:1604.01706(2016).
34. L. Zhou, K. Xu, A. Zubair, A. D. Liao, W. Fang, F. Ouyang, Y. H. Lee, K. Ueno, R. Saito, T. Palaciios, J. Kong, and M. S. Dresselhaus: *Large-area synthesis of high-quality uniform few-layer MoTe$_2$*. Journal of the American Chemical Society **137**(37), 11892(2015).
35. J. Zhou, F. Liu, J. Lin, X. Huang, J. Xia, B. Zhang, Q. Zheng, H. Wang, C. Zhu, L. Niu, X. Wang, W. Fu, P. Yu, T-R. Chang, C-H. Hsu, D. Wu, H-T. Jeng, Y. Huang, H. Lin, Z. Shen, C. Yang, L. Lu, K. Suenaga, W. Zhou, S. T. Pantelides, G. Liu, and Z. Liu: *Large-area and high-quality 2D transition metal telluride*. Advanced Materials **29**(3), 1603471(2017).
36. B. E. Brown: *The crystal structures of WTe$_2$ and high-temperature MoTe$_2$*. Acta Crystallographica **20**(2), 268(1966).
37. D. Shin, Y. Lee, M. Sasaki, Y. H. Jeong, F. Weickert, J. B. Betts, H. J. Kim, K. S. Kim, and J. Kim: *Violation of Ohm's law in a Weyl metal*. Nature materials **16**(11), 1096(2017).
38. F. C. Chen, H. Y. Lv, X. Luo, W. J. Lu, Q. L. Pei, G. T. Lim, Y. Y. Han, X. B. Zhu, W. H. Song, and Y. P. Sun: *Extremely large magnetoresistance in the type-II Weyl semimetal Mo Te$_2$*. Physical Review B **94**(23), 235154(2016).
39. M. N. Ali, J. Xiong, S. Flynn, J. Tao, Q. D. Gibson, L. M. Schoop, T. Liang, N. Haldolaarachchige, M. Harschberger, N. P. Ong, and J. Cava: *Large, non-saturating magnetoresistance in WTe$_2$*. Nature **514**(7521), 205(2014).
40. L. X. Yang, Z. K. Liu, Y. Sun, H. Peng, H. F. Yang, T. Zhang, B. Zhou, Y. Zhang, Y. F. Guo, M. Rahn, D. Prabhakaran, Z. Hussain, S.K. Mo, C. Felser, B. Yan, and Y. L. Chen: *Weyl semimetal phase in the non-centrosymmetric compound TaAs*. Nature physics **11**(9), 728(2015).
41. C. Xu, B. Li, W. Jiao, W. Zhou, B. Qian, R. Sankar, N. D. Zhigadlo, Y. Qi, D. Qian, F. C. Chou, and X. Xu: *Topological type-II Dirac fermions approaching the Fermi level in a transition metal dichalcogenide NiTe$_2$*. Chemistry of materials **30**(14), 4823(2018).
42. D. D. Liang, Y. J. Wang, W. L. Zhen, J. Yang, S. R. Weng, X. Yan, Y. Y. Han, W. Tong, W. K. Zhu, L. Pi, and C. J. Zhang: *Origin of planar Hall effect in type-II Weyl semimetal MoTe$_2$*. AIP Advances **9**(5) 055015(2019).
43. J.-X. Gong, J. Yang, M. Ge, Y.-J. Wang, D.-D. Liang, L. Luo, X. Yan, W.-L. Zhen, S.-R. Weng, L. Pi, C.-J. Zhang, and W.-K. Zhu: *Non-stoichiometry effects on the extreme magnetoresistance in Weyl semimetal WTe$_2$*. Chinese Physics Letters **35**(9) 097101(2018).
44. A. W. Forbes, R. P. Dulal, N. Bhattarai, I. L. Pegg, and J. Philip: *Experimental realization and magnetotransport properties of half-metallic Fe$_2$Si*. Journal of Applied Physics **125**(24), 243902(2019).
45. A. J. Bestwick, E. J. Fox, X. Kou, L. Pan, K. L. Wang, and D. G-Gordon: *Precise quantization of the anomalous Hall effect near zero magnetic field*. Physical review letters **114**(18), 187201(2015).
46. C.-Z. Chang, J. Zhang, X. Feng, J. Shen, Z. Zhang, M. Guo, K. Li, Y. Ou, P. Wei, L.-L. Wang, Z.-Q. Ji, Y. Feng, S. Ji, X. Chen, J. Jia, X. Dai, Z. Fang, S.-C. Zhang, K. He, Y. Wang, L. Lu, X.-C. Ma, and Q.-K. Xue: *Experimental observation of the quantum anomalous Hall effect in a magnetic topological insulator*. Science **340**(6129), 167(2013).